\documentclass[aps,prl,twocolumn,showpacs,superscriptaddress]{revtex4}

\usepackage{mathrsfs}
\usepackage{epsfig}
\usepackage{graphicx}
\usepackage{amsfonts}
\usepackage[figuresright]{rotating}
\usepackage{amssymb}
\usepackage{amsmath}
\usepackage{dcolumn}
\usepackage{bm}

\def\be{\begin{equation}}
\def\ee{\end{equation}}
\def\bea{\begin{eqnarray}}
\def\eea{\end{eqnarray}}

\begin{document}

\title{Phase-Sensitive Detection for Unconventional Bose-Einstein
Condensations}
\author{Zi Cai}
\affiliation{Department of Physics, University of California, San
Diego, California 92093}
\author{Lu-Ming Duan}
\affiliation{Department of Physics and MCTP, University of Michigan,
Ann Arbor, Michigan 48109, USA\\
and Center for Quantum Information, IIIS, Tsinghua University, Beijing, China}
\author{Congjun Wu}
\affiliation{Department of Physics, University of California, San
Diego, California 92093}

\begin{abstract}
We propose a phase-sensitive detection scheme to identify the unconventional
$p_{x}\pm ip_{y}$ symmetry of the condensate wavefunctions of bosons, which
have already been proposed and realized in high bands in optical lattices.
Using the impulsive Raman operation combining with time-of-flight imaging,
the off-diagonal correlation functions in momentum space give rise to the
relative phase information between different components of condensate
wavefunctions.
This scheme is robust against the interaction and interband
effects, and provides smoking gun evidence for unconventional
Bose-Einstein condensations with nontrivial condensation symmetries.
\end{abstract}
\pacs{67.85.Jk, 67.85.Hj, 05.30.Jp}
\maketitle

Macroscopic condensates of bosons and paired fermions are of central interest
in condensed matter physics. Order parameters of Cooper pairings are
termed \textquotedblleft unconventional\textquotedblright\ if they belong
to non-trivial representations of rotational symmetry groups.
Celebrated examples include the $d$-wave pairing states of high $T_{c}$
cuprates \cite{wollman1993,tsuei1994}, the $p$-wave pairing states of
$^{3}$He $A$ and $B$-phases \cite{leggett1975}, and Sr$_{2}$RuO$_{4}$
\cite{nelson2004,kidwingira2006}. Among various experimental tools,
phase-sensitive detections are exceptional as they provide the smoking-gun
evidence for unconventional pairing symmetries, such as the $\pi $-phase
shifts in the joint corner SQUID junctions \cite{wollman1993} and the
tri-crystal superconducting ring experiments \cite{tsuei1994} of high $T_{c}$
cuprates. Unconventional symmetries have also been generalized to the
particle-hole channel pairings, \textit{i.e.}, the Pomeranchuk type Fermi
surface instabilities in both density \cite{fradkin2011} and spin channels
\cite{wu2004,wu2007a}. The spin instabilities in high orbital angular
momentum channels are denoted as \textquotedblleft unconventional
magnetism\textquotedblright.

Recently, unconventional symmetries have been introduced to the single-boson
condensates \cite{wu2009,cai2011}, denoted as ``unconventional''
Bose-Einstein condensations (UBEC). Their condensate wavefunctions belong to
non-trivial representations of the lattice point group. Such states have
been proposed in high orbital bands of optical lattices \cite%
{isacsson2005,liu2006,wu2006,kuklov2006,stojanovic2008,zhou2011}. These
systems have been experimentally realized by pumping bosons into high
orbital bands \cite{mueller2007,wirth2011,wirth2011a}. Bosons have been
observed to develop phase coherence before they decay to the lowest band.
In addition, in the artificial lattice systems of exciton-polariton
in semiconductor quantum wells,  a $d$-wave
condensation in the excited bands of bosons have also been observed \cite%
{kim2011}. These UBECs are beyond Feynman's ``no-node'' theorem \cite%
{wu2009,feynman1972}, which states that the ground-state wavefunctions of
bosons are positive-definite under very general circumstances. This theorem
applies for the system of superfluid $^4$He \cite{feynman1972} and
many experiments of alkali bosons \cite{pitaevskii1999}. It also implies
that time-reversal (TR) symmetry cannot be spontaneously broken in usual
BECs. However, UBECs escape from the ``no-node'' constraint. Their
condensate wavefunctions are nodal, which are able to break TR symmetry
spontaneously under certain conditions \cite{wu2009}.

The recent UBECs realized in Hemmerich's group is an exciting progress \cite%
{wirth2011}, where the time-of-flight (TOF) spectrum has revealed signatures
of both the real and complex UBECs by tuning the anisotropy of the optical
lattice. However, the TOF images can only provide the single-particle
density distribution in momentum space, thus in the complex UBECs, the key
information about the relative phase between the condensate components is
lost during TOF. Without the phase information, the TOF images of the $%
p_{x}\pm ip_{y}$ BEC can be interpreted by other plausible scenarios such as
the phase separation between two real condensates at different momenta or
the incoherent mixing between them. It would be important to have the
smoking gun evidence of the phase difference $\pm \frac{\pi}{2}$ between
the two condensate components.

In this paper, we propose a phase-sensitive detection scheme to identify the
$p_{x}\pm ip_{y}$ symmetry of UBECs by measuring the relative phases of $\pm
\frac{\pi}{2}$.
This proposal is based on the scheme in Ref. \cite{duan2006}, which
has been used to construct the off-diagonal correlation functions in momentum
space for UBECs. By implementing a momentum-kick Raman
pulse, we build up the connection between bosons with different condensate
momenta in the complex UBEC. As we will show below, the relative phase
information is uniquely tied to the off-diagonal correlations between the
different condensate momenta, which can be measured in time-of-flight
imaging through the impulsive Raman pulse. We note that a different scheme
has been proposed recently by Kitagawa $et$ $al$. for phase-sensitive
detection of nontrivial pairing symmetries in ultracold fermions based on
the two-particle interferometry \cite{kitagawa2011}.

\begin{figure}[htb]
\includegraphics[width=0.98\linewidth]{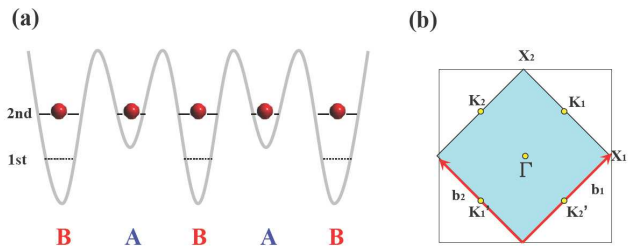}
\caption{(a) The bipartite lattice structure (along the
$x$-direction) in Hemmerich's experiment \protect\cite{wirth2011},
where bosons are loaded into the second band. (b) The first
Brillouin zone (blue area) and the basis vectors of the reciprocal
lattice $\vec b_{1,2}$.} \label{fig:lattice}
\end{figure}

Our scheme is connected with the experiment in Ref.
\cite{wirth2011}, where the bosons are pumped to the first excited
band of a $s$-$p$ hybridized system (hybridization between the
$s$-orbital of the shallow sublattice $A$ and the $p$-orbital of the
deep sublattice $B$ as illustrated in Fig. \ref{fig:lattice}(a). Two
degenerate band minima (denoted as $K_{1,2}$ below) locate at the
lattice momenta $\frac{1}{2}\vec{b}_{1,2}$ with $\vec{b}_{1,2}=(\pm
\pi ,\pi )$ the reciprocal lattice vectors (we set the lattice
constant $a=1$), as shown in Fig.\ref{fig:lattice} (b). Bloch-wave
states $\Psi _{1,2}(\mathbf{r})$ at $K_{1,2}$ points are real-valued
wavefunctions with nodal lines similar to standing waves, and thus
are time-reversal invariant.
The complex combination of $\Psi _{\pm }=(\Psi _{1}\pm i\Psi _{2})/%
\sqrt{2}$ only have nodal points from intersections of the nodal lines of $%
\Psi _{1,2}$.
With repulsive interaction, the complex condensates $\Psi _{\pm
}$ with nodal points are favored since their spatial distributions of
particle density are more uniform and extensive than other states,
minimizing the interaction energy \cite{cai2011}.

Both of the wavefunctions $\Psi _{1,2}(\mathbf{r})$ have odd parity
with the $p$-wave symmetry. Rigorously speaking, the lattice
configuration in Ref. \cite{wirth2011} does not have 4-fold
rotational symmetry, and thus $\Psi_{1,2}$ are not transformable to
each other by the rotation of 90$^\circ$. Nevertheless, for
simplicity, we still denote these condensates with the $p_x\pm i
p_y$ symmetry. The TOF imaging in the experiment \cite{wirth2011}
has observed four peaks at ($K_{1,2}$, $K_{1,2}^{\prime }\equiv
K_{1,2}-\vec{b}_{1,2}$) with the same height, which implies that the
condensate has equal weights of the components $\Psi _{1}$ and $\Psi
_{2}$. However, the phase difference between the two components
$\Psi_1$ and $\Psi_2$, which is critical for verifying the novel
$p_x \pm i p_y$ condensation symmetry, is not clear.

\begin{figure}[tbh]
\includegraphics[width=0.85\linewidth]{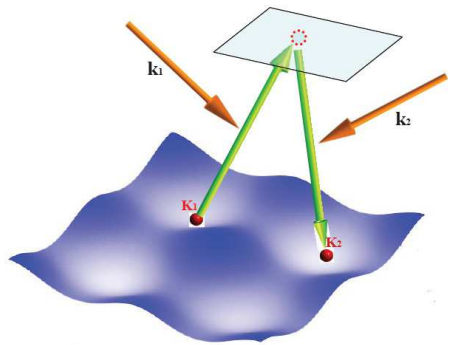}
\caption{Using two Raman pulses with different propagating directions to
built up the momentum transfer between bosons at $K_{1}$ and $K_{2}$.}
\label{fig:Raman}
\end{figure}

The spirit of our proposal for the phase-sensitive detection
can be outlined of as follows: for the condensate
$|\Psi_c \rangle =\frac{1}{2^{N_c/2}\sqrt{N_c!}}%
(C_{K_{1}}^{\dag }+e^{i\theta }C_{K_{2}}^{\dag })^{N_c}|O\rangle $,
where $N_c$ is the boson number in the condensate; $C_{K_{1}}^{\dag
}$ ($C_{K_{2}}^{\dag }$) is the bosonic operator creating a boson at
$K_{1}$ ($K_{2}$); $\theta $ is the relative phase between bosons in
$K_{1}$ and $K_{2}$. If $\theta\neq 0$ and $\pi$, the condensate
exhibits a vortex-antivortex lattice structure. The staggered
orbital angular momentum (OAM) density wave order parameter is
defined $L_z(\vec Q)=i(C_{K_{1}}^{\dag }C_{K_{2}}-C_{K_{2}}^{\dag
}C_{K_{1}})$ where $\vec Q=\vec K_1-\vec K_2$. Its magnitude reads
as $ \langle \Psi_c | L_z(\vec Q) |\Psi_c \rangle = N_c \sin
\theta$, which is just the off-diagonal correlation.
It  reaches maximum for the $p_{x}\pm ip_{y}$ state ($%
\theta =\pm \frac{1}{2}\pi$).
If we implement a Raman
transition to transform the bosons in the original condensate into \cite%
{duan2006}: \bea C_{K_{1}}^{\prime }=\frac{1}{\sqrt
2}(C_{K_{1}}-iC_{K_{2}}); ~~ C_{K_{2}}^{\prime }=\frac{1}{\sqrt
2}(C_{K_{2}}-iC_{K_{1}}), \label{eq:transformation} \eea we can
measure the density difference of the new BECs $\delta n^{\prime
}\equiv \langle \Psi_C| C_{K_{2}}^{\prime \dag }C_{K_{2}}^{\prime
}-C_{K_{1}}^{\prime \dag }C_{K_{1}}^{\prime } |\Psi_c \rangle $
through the TOF imaging. $\delta n^{\prime }$ exactly gives the
desired off-diagonal correlation as \bea \delta n^{\prime }=i\langle
\Psi_c|C_{K_{1}}^{\dag }C_{K_{2}}-C_{K_{2}}^{\dag }C_{K_{1}} |\Psi_c
\rangle. \eea

Now we turn back to the Hemmerich's experiment and show how to
implement the Raman transition. Similar to Ref. \cite{duan2006}, the
Raman transition can be realized by two traveling-wave laser beams
propagating along different
directions (as plotted in Fig.\ref{fig:Raman}) with corresponding wavevector $%
\mathbf{k_{1,2}}$ and frequency $\omega _{1,2}$, which introduce an
effective Raman Rabi frequency with a spatially varying phase $
\Omega (\mathbf{r},t) = \Omega _{0}e^{i(\delta \mathbf{k}\cdot \mathbf{r-\delta \omega t}%
+\phi )}$, where $\delta \mathbf{k=k_{1}-k_{2}}$, $\delta \omega \mathbf{=}%
\omega _{1}-\omega _{2}$, and $\phi $ is the relative phase between
the two Raman beams. $\Omega_0$ is expressed as
$\Omega_0=\Omega_1\Omega_2^*/\Delta$, where $\Delta$ is the
detuning, $\Omega_{1(2)}$ are the resonant Rabi frequencies for the
individual transitions between the initial (finial) states and the
intermediate state, and are proportional to the strength of the
electric field of the corresponding Raman beams. This spatially
dependent Raman transition builds up the connection between the
condensation components in two degenerate points $K_{1,2}=(\pm \pi
/2,\pi /2)$,
which demands that $\delta \mathbf{k}=K_{2}-K_{1}=(\pi ,0)$ (Notice that $%
\delta \mathbf{k}$ and $-\delta \mathbf{k}$ are connected by a
reciprocal lattice vector and thus equivalent. Because of this
feature, the Raman scheme here is simpler compared with the one in
Ref. \cite{duan2006} which needs to use transitions between two
different hyperfine levels). The effective Hamiltonian for the Raman
process is described by
\begin{equation}
H_{R}=\int d\mathbf{r}\Omega (\mathbf{r},t)\Psi ^{\dag }(\mathbf{r})
\Psi (\mathbf{r})+h.c, \label{Eq:Rabi1}
\end{equation}
which, together with the original atomic Hamiltonian $H_{0}$ in the
optical lattice\cite{wirth2011}, gives the full Hamiltonian of the
system.

\begin{figure}[htb]
\includegraphics[width=0.8\linewidth]{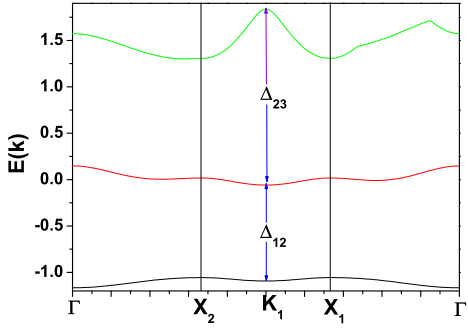}
\caption{The energy spectra along the lines between the high
symmetry points with the experimental values of the parameters given
in Ref.\protect\cite{wirth2011}.} \label{fig:energyband}
\end{figure}

Different from Ref.\cite{duan2006}, in the experiment
\cite{wirth2011} the optical lattice is too shallow to apply the
tight binding approximation. Instead, we expand the field operator
in the Bloch representation as
\begin{equation}
\Psi (\mathbf{r})=\sum_{n\mathbf{k}}C_{n\mathbf{k}}\psi _{n\mathbf{k}}(%
\mathbf{r}),  \label{Eq:Bloch}
\end{equation}%
where $C_{n\mathbf{k}}$ is the bosonic operator annihilating a boson in the $%
n$th band with momentum $\mathbf{k}$, $\psi _{n\mathbf{k}}(\mathbf{r})$ is
the Bloch wavefunction, and the summation of $\mathbf{k}$ is over the first
BZ. We choose the effective Rabi frequency $\Omega _{0}$ of the Raman pulse
so that it is small compared with the band gap but large compared with the
atomic hopping rate in the lattice ($t\ll \hbar \Omega _{0}\ll \Delta _{12}$%
).  Under this condition, we can neglect the interband tunneling as well as
the time-dependence of the wavepackets $\Psi _{1}(\mathbf{r})$ and $\Psi _{2}(%
\mathbf{r})$ during the Raman transition. For the typical values of
the experimental parameters, the energy band structure is shown in Fig. \ref%
{fig:energyband}. The hopping rate is estimated by $t\approx
0.05E_{r}\approx 2\pi \hbar \times 0.1$ kHz and the smallest bandgap
$\Delta _{12}\approx 1.08E_{r}\approx 2\pi \hbar \times 2.2$ kHz. If
we choose $\hbar \Omega _{0}\sim 2\pi\hbar \times 0.5$ kHz, the
corrections to the above approximation, estimated by $t^{2}/\left(
\hbar \Omega _{0}\right) ^{2}$ and $\left( \hbar \Omega _{0}\right)
^{2}/\Delta _{12}^{2}$, are pretty small. In our case, the Raman
operation induces a transition between the complex and polar UBECs
with the same kinetic energy but different interaction energy.
$\hbar\delta \omega$ should match the energy difference between the
initial (complex UBEC) and final (real UBEC) states of the Raman
transition, which can be estimated as $10^{-3}E_r\approx 2\pi \hbar
\times 2.1$ Hz and much smaller than $\hbar \Omega _{0}$. Therefore,
the phase accumulation induced by $\delta \omega$ within the
duration of Raman pulses $\delta t$ can be neglected ($\delta
\omega\delta t\ll \pi/4$), and the Raman Rabi frequency in
Eq.(\ref{Eq:Rabi1}) can be considered as time-independent during the
Raman transition.

\begin{figure}[tbh]
\includegraphics[width=0.95\linewidth]{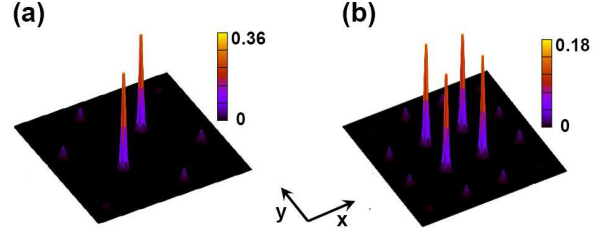}
\caption{TOF imagings after the Raman transition  with $\protect\phi
=0$ for (a) the complex UBEC ($\Psi_1+i\Psi_2$) and (b) incoherent
mixing of the real UBECs ($\Psi_1$ and $\Psi_2$). } \label{fig:TTT}
\end{figure}

Under the above approximations, $H_{R}$ is simplified to
\begin{equation}
H_{R}=e^{i\phi }\sum_{\mathbf{k}}\Omega
(\mathbf{k})C_{\mathbf{k+}\delta \mathbf{k}}^{\dag
}C_{\mathbf{k}}+h.c,  \label{Eq:RO1}
\end{equation}%
which is constrained only to the relevant band. The summation of
$\mathbf{k}$ is
over the first BZ, and the effective Raman-Rabi frequency $\Omega (\mathbf{k}%
)$ is $\mathbf{k}$ dependent and can be calculated based on the
eigenvectors obtained in the band-structure calculation. Before the
Raman transition, the momentum distribution of the $p_{x}\pm ip_{y}$
condensate is sharply peaked at $K_{1}$ and $K_{2}$ with a small
distribution width $\Lambda$, and within this small region $\Omega
(\mathbf{k})$ can be considered as a constant, estimated by $\Omega
\approx 0.98\Omega _{0}$ for the typical values of the experimental
parameters. So during the Raman transition, the small wave-packets
around $K_{1}$ and $K_{2}$ are transferred by the same formula
without distortion. When we choose the duration of the Raman pulse
as $\Omega \delta t=\pi /4$, the wave-packets around $K_{1}$ and
$K_{2}$ with $\left\vert d\mathbf{k}\right\vert\leq \Lambda $ are
transferred by
\begin{eqnarray}
C_{\mathbf{K}_{1}+d\mathbf{k}}^{\prime } &=&(C_{\mathbf{K}_{1}+d\mathbf{k}%
}-ie^{i\phi }C_{\mathbf{K}_{2}+d\mathbf{k}})/\sqrt{2},  \notag \\
C_{\mathbf{K}_{2}+d\mathbf{k}}^{\prime } &=&(C_{\mathbf{K}_{2}+d\mathbf{k}%
}-ie^{-i\phi }C_{\mathbf{K}_{1}+d\mathbf{k}})/\sqrt{2}.  \label{simple}
\end{eqnarray}

Right after this Raman transition, we turn off the trapping optical
potential and perform the TOF imaging to measure the particle
density distribution in momentum space around $\mathbf{K}_{1(2)}$.
The density difference  $\langle
\delta n^{\prime }\left( d\mathbf{k}\right) \rangle \equiv \langle n_{K_{2}+d%
\mathbf{k}}^{\prime }-n_{K_{1}+d\mathbf{k}}^{\prime }\rangle
=\langle C_{K_{2}+d\mathbf{k}}^{\prime \dag
}C_{K_{2}+d\mathbf{k}}^{\prime }-C_{K_{1}+d\mathbf{k}}^{\prime \dag
}C_{K_{1}+d\mathbf{k}}^{\prime }\rangle $ gives the off diagonal
correlation for the original UBEC
\begin{equation}
\delta n^{\prime }\left( d\mathbf{k}\right) =i\langle e^{i\phi }C_{K_{1}+d%
\mathbf{k}}^{\dag }C_{K_{2}+d\mathbf{k}}-e^{-i\phi }C_{K_{2}+d\mathbf{k}%
}^{\dag }C_{K_{1}+d\mathbf{k}}\rangle.\label{Eq:deltan}
\end{equation}%
Notice that the experimental observable $\delta n^{\prime }\left(
d\mathbf{k=}0\right)$, the height difference between  the peaks in
$K_1$ and $K_2$, is dependent on the phase difference of the two
Raman pulses $\phi$. For the $p_x \pm ip_y$ BEC, after the Raman
pulse, we see that $\delta n^{\prime}\left(
d\mathbf{k=}0\right)\propto\cos\phi$ from Eq.(\ref{Eq:deltan}). The
oscillation of $\delta n^{\prime}$ with $\phi$ indicates coherence
of the two Raman pulses, which is critical for phase-sensitive
detection. For $\phi=0$, $\delta n^{\prime }\left(
d\mathbf{k=}0\right) =i\langle C_{K_{1}}^{\dag
}C_{K_{2}}-C_{K_{2}}^{\dag }C_{K_{1}}\rangle$ represents the order
parameter of the orbital ordering of the original UBEC. With this
phase-sensitive measurement, we can easily distinguish the complex
condensate $|\Psi \rangle \propto
\frac{1}{2^{N_c/2}\sqrt{N_c!}}(C_{K_{1}}^{\dag }+e^{i\theta
}C_{K_{2}}^{\dag })^{N_c}|O\rangle $ and other plausible scenarios,
such as the phase separation or incoherent mixing between the polar
UBECs $|\Psi _{1}\rangle =\frac{1}{\sqrt{N_c!}}(C_{K_{1}}^{\dag
})^{N_c}|O\rangle $ and $|\Psi _{2}\rangle =\frac{1}{\sqrt{N_c!}}%
(C_{K_{2}}^{\dag })^{N_c}|O\rangle $. In the conventional TOF
imagings, both of them exhibit four peaks with the same height thus
can not be distinguished. Under this phase-sensitive TOF\
imaging, we find that for the $p_{x}\pm ip_{y}$ UBEC, $\langle \Psi |n_{K_{2}+d%
\mathbf{k}}^{\prime }|\Psi \rangle =0$, which means that after Raman
transition, the initial complex UBEC turns to the polar UBEC
condensing only at $K_1$, and the predicted new TOF images are shown
in Fig.\ref{fig:TTT} (a) with only two peaks, in contrast with the
four peaks that one expects to see for the incoherent mixing state
between the condensates $\Psi _{1}$ and $\Psi _{2}$ show in Fig.
\ref{fig:TTT} (b).

\begin{figure}[tbh]
\includegraphics[width=0.97\linewidth]{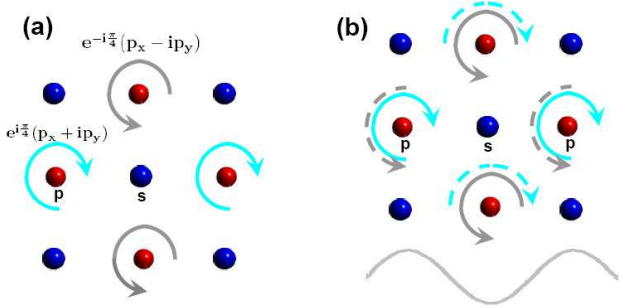}
\caption{(a) The vortex-antivortex lattice pattern of the $p_x\pm
ip_y$ BEC before the Raman transition; (b) the real-space current
pattern during the Raman transition, solid (dashed) arrows denote
original (reflected) currents.} \label{fig:real}
\end{figure}

To get a better understanding of our results, we provide a real
space picture to illustrate the complex-real UBEC transition during
the Raman process. Before the Raman transition, the $p_x\pm ip_y$
UBEC exhibits a vortex-antivortex lattice structure in sublattice
$B$ ($p$-orbital sites), as shown in Fig.\ref{fig:real}(a). The
Raman beams introduce an extra potential with the form of
Eq.(\ref{Eq:Rabi1}), as shown in Fig.\ref{fig:real} (b). Without
loss of generality, we focus on one site in sublattice $B$,
initially the local wavefunction within this site can be
approximately considered as $\varphi(\mathbf{r})\sim
e^{i\frac{\pi}4}[\varphi_x(\mathbf{r})+i\varphi_y(\mathbf{r})]$,
where $\varphi_{x(y)}(\mathbf{r})$ denote the $p_{x(y)}$-orbital
Wannier function. During the Raman transition, the initial current
is reflected by the extra potential, and the local wavefunction
turns to $\varphi(t)=\cos(\Omega t)\varphi(\mathbf{r})+\sin(\Omega
t)\varphi^*(\mathbf{r})$, where $\varphi^*(\mathbf{r})$ is the TR
counterpart of $\varphi(\mathbf{r})$ carrying a current with an
opposite direction. Initially, $\varphi(0)=\varphi(\mathbf{r})$
denotes the $p_x+ip_y$ state, at the momentum of $t_0\Omega=\pi/4$,
the reflected current happens to cancel with the initial one, and it
turns to a polar state with the real wavefunction $\varphi(t_0)\sim
\varphi_x(\mathbf{r})-\varphi_y(\mathbf{r})$, and the corresponding
polar UBEC exhibits two peaks in the TOF spectrum, as shown in
Fig.\ref{fig:TTT} (a).

\begin{figure}[tbh]
\includegraphics[width=0.8\linewidth]{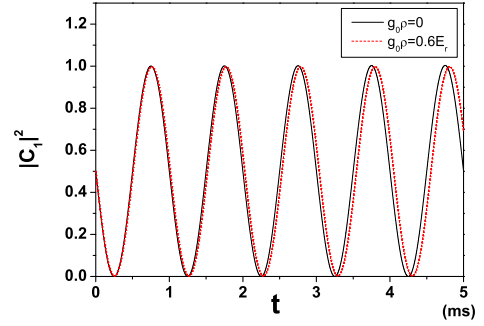} .
\caption{ Time evolution of the condensation fraction at $\Psi_{K_{1}}$ with Raman frequency $\Omega =2%
\protect\pi \times 0.5$kHz, and interaction parameter
$g_{0}\protect\rho =0$ (solid line) and $0.6E_{r} $ (dashed line),
respectively} \label{fig:Rabi}
\end{figure}

In above analysis of the Raman transition, we have neglected the
effect of interaction, which is responsible for the broadening of
the TOF imaging peaks \cite{yi2007,lin2008}. Now we estimate the
interaction effect by solving the time evolution from the
Gross-Pitaevskii (GP) equations (the GP\ equation gives an adequate
description of the interaction since the initial state of the system
is a BEC). A similar problem has been addressed for the many-body
Rabi oscillation in a two-component BEC \cite{saito2007}. As
analyzed above, we neglect the deformation of $\Psi _{1}(\mathbf{r})$ and $%
\Psi _{2}(\mathbf{r})$ during the time evolution, and the dynamics
of the system can be approximately considered as two-mode
transition, thus the wavefunction during the time evolution can be
expressed as: $\Psi (t)=C_{1}(t)\Psi _{1}+C_{2}(t)\Psi _{2}$.  The
corresponding GP equation reads:
\begin{equation}
i\frac{\partial \Psi }{\partial t}=\Big\{H_{0}+\Omega (r)+g_{0}\rho |\Psi (%
\vec{r})|^{2}\Big\}\Psi (\vec{r}),  \label{GP}
\end{equation}%
where $H_{0}=-\frac{\hbar ^{2}\vec{\nabla}^{2}}{2M}+V_{0}(\vec{r})$ is the
original optical lattice Hamiltonian, $\rho $ is the average density, $g_{0}$
is the $s$-wave scattering interaction parameter. In the experiment\cite%
{wirth2011}, $g_{0}\rho $ is estimated to be $0.6E_{r}$. Notice that $\psi
_{1(2)}$ are eigenfunctions of $H_{0}$. thus we get:
\begin{eqnarray}
i\frac{\partial C_{1}(t)}{\partial t} &=&\Omega
C_{2}+(2g|C_{1}|^{2}+4g^{\prime }|C_{2}|^{2})C_{1}+2g^{\prime }C_{1}^{\ast
}C_{2}^{2},  \notag \\
i\frac{\partial C_{2}(t)}{\partial t} &=&\Omega C_{1}+(4g^{\prime
}|C_{2}|^{2}+2g|C_{2}|^{2})C_{2}+2g^{\prime }C_{2}^{\ast }C_{1}^{2},
\notag \label{GP2}
\end{eqnarray}%
where $g=g_{0}\rho \int d^{2}r|\Psi _{1}(r)|^{4}=g_{0}\rho \int d^{2}r|\Psi
_{2}(r)|^{4}$, $g^{\prime}=g_{0}\rho \int d^{2}r|\Psi _{1}(r)|^{2}|\Psi _{2}(r)|^{2}$.%
 Using the initial condition: $C_{1}(0)=1/\sqrt{2}$, $C_{2}(0)=i/\sqrt{2}$ (%
$\Psi_{K_1}+i\Psi_{K_2}$ state) and $\Omega =2\pi \times 0.5$ kHz,
we obtain the time evolution of condensation fraction of bosons at
$K_{1}$ wave-packet, and compare the result with the non-interacting
case. As shown in Fig. \ref{fig:Rabi}, the interaction barely
changes the Rabi oscillation within the duration of the Raman pulses
in our case, which implies the single-particle Rabi oscillation
approximation we adopted above provides an accurate description for
the many-body dynamics of the system during the Raman transition.

To conclude, we propose a phase-sensitive detection scheme to
identify the nontrivial symmetry of recently observed $p_{x}\pm
ip_{y}$ orbital UBEC, where the connection between different
condensate components is built up by the momentum kick provided by
impulsive Raman pulses. Our scheme can also be applied to the
phase-sensitive detections for unconventional BECs with other
symmetries, $\mathit{eg.}$ the recently observed UBEC in $f$-orbital
bands of the optical lattice \cite{wirth2011a}.

This work was supported by the NSF DMR-1105945, the AFOSR-YIP program, the
NBRPC (973 Program) 2011CBA00300 (2011CBA00302), the DARPA OLE program, the
IARPA MUSIQC program, the ARO and the AFOSR MURI program.


\begin{thebibliography}{27}
\expandafter\ifx\csname
natexlab\endcsname\relax\def\natexlab#1{#1}\fi
\expandafter\ifx\csname bibnamefont\endcsname\relax
  \def\bibnamefont#1{#1}\fi
\expandafter\ifx\csname bibfnamefont\endcsname\relax
  \def\bibfnamefont#1{#1}\fi
\expandafter\ifx\csname citenamefont\endcsname\relax
  \def\citenamefont#1{#1}\fi
\expandafter\ifx\csname url\endcsname\relax
  \def\url#1{\texttt{#1}}\fi
\expandafter\ifx\csname urlprefix\endcsname\relax\def\urlprefix{URL
}\fi \providecommand{\bibinfo}[2]{#2}
\providecommand{\eprint}[2][]{\url{#2}}

\bibitem[{\citenamefont{Wollman et~al.}(1993)\citenamefont{Wollman,
  Van~Harlingen, Lee, Ginsberg, and Leggett}}]{wollman1993}
\bibinfo{author}{\bibfnamefont{D.~A.} \bibnamefont{Wollman}},
  \bibinfo{author}{\bibfnamefont{D.~J.} \bibnamefont{Van~Harlingen}},
  \bibinfo{author}{\bibfnamefont{W.~C.} \bibnamefont{Lee}},
  \bibinfo{author}{\bibfnamefont{D.~M.} \bibnamefont{Ginsberg}},
  \bibnamefont{and} \bibinfo{author}{\bibfnamefont{A.~J.}
  \bibnamefont{Leggett}}, \bibinfo{journal}{Phys. Rev. Lett.}
  \textbf{\bibinfo{volume}{71}}, \bibinfo{pages}{2134} (\bibinfo{year}{1993}).

\bibitem[{\citenamefont{Tsuei et~al.}(1994)\citenamefont{Tsuei, Kirtley, Chi,
  Yu-Jahnes, Gupta, Shaw, Sun, and Ketchen}}]{tsuei1994}
\bibinfo{author}{\bibfnamefont{C.~C.} \bibnamefont{Tsuei}},
  \bibinfo{author}{\bibfnamefont{J.~R.} \bibnamefont{Kirtley}},
  \bibinfo{author}{\bibfnamefont{C.~C.} \bibnamefont{Chi}},
  \bibinfo{author}{\bibfnamefont{L.~S.} \bibnamefont{Yu-Jahnes}},
  \bibinfo{author}{\bibfnamefont{A.}~\bibnamefont{Gupta}},
  \bibinfo{author}{\bibfnamefont{T.}~\bibnamefont{Shaw}},
  \bibinfo{author}{\bibfnamefont{J.~Z.} \bibnamefont{Sun}}, \bibnamefont{and}
  \bibinfo{author}{\bibfnamefont{M.~B.} \bibnamefont{Ketchen}},
  \bibinfo{journal}{Phys. Rev. Lett.} \textbf{\bibinfo{volume}{73}},
  \bibinfo{pages}{593} (\bibinfo{year}{1994}).

\bibitem[{\citenamefont{Leggett}(1975)}]{leggett1975}
\bibinfo{author}{\bibfnamefont{A.~J.} \bibnamefont{Leggett}},
  \bibinfo{journal}{Rev. Mod. Phys.} \textbf{\bibinfo{volume}{47}},
  \bibinfo{pages}{331} (\bibinfo{year}{1975}).

\bibitem[{\citenamefont{Nelson et~al.}(2004)\citenamefont{Nelson, Mao, Maeno,
  and Liu}}]{nelson2004}
\bibinfo{author}{\bibfnamefont{K.}~\bibnamefont{Nelson}},
  \bibinfo{author}{\bibfnamefont{Z.}~\bibnamefont{Mao}},
  \bibinfo{author}{\bibfnamefont{Y.}~\bibnamefont{Maeno}}, \bibnamefont{and}
  \bibinfo{author}{\bibfnamefont{Y.}~\bibnamefont{Liu}},
  \bibinfo{journal}{Science} \textbf{\bibinfo{volume}{12}},
  \bibinfo{pages}{1151} (\bibinfo{year}{2004}).

\bibitem[{\citenamefont{Kidwingira et~al.}(2006)\citenamefont{Kidwingira,
  Strand, Harlingen, and Maeno}}]{kidwingira2006}
\bibinfo{author}{\bibfnamefont{F.}~\bibnamefont{Kidwingira}},
  \bibinfo{author}{\bibfnamefont{J.~D.} \bibnamefont{Strand}},
  \bibinfo{author}{\bibfnamefont{D.~J.~V.} \bibnamefont{Harlingen}},
  \bibnamefont{and} \bibinfo{author}{\bibfnamefont{Y.}~\bibnamefont{Maeno}},
  \bibinfo{journal}{Science} \textbf{\bibinfo{volume}{314}},
  \bibinfo{pages}{1267} (\bibinfo{year}{2006}).

\bibitem[{\citenamefont{Fradkin}(2009)}]{fradkin2011}
\bibinfo{author}{\bibfnamefont{E.}~\bibnamefont{Fradkin}},
  \emph{\bibinfo{title}{Lectures at the Les Houches Summer School on "Modern
  theories of correlated electron systems"}} (\bibinfo{year}{2009}).

\bibitem[{\citenamefont{Wu and Zhang}(2004)}]{wu2004}
\bibinfo{author}{\bibfnamefont{C.}~\bibnamefont{Wu}} \bibnamefont{and}
  \bibinfo{author}{\bibfnamefont{S.-C.} \bibnamefont{Zhang}},
  \bibinfo{journal}{Phys. Rev. Lett.} \textbf{\bibinfo{volume}{93}},
  \bibinfo{pages}{036403} (\bibinfo{year}{2004}).

\bibitem[{\citenamefont{Wu et~al.}(2007)\citenamefont{Wu, Sun, Fradkin, and
  Zhang}}]{wu2007a}
\bibinfo{author}{\bibfnamefont{C.}~\bibnamefont{Wu}},
  \bibinfo{author}{\bibfnamefont{K.}~\bibnamefont{Sun}},
  \bibinfo{author}{\bibfnamefont{E.}~\bibnamefont{Fradkin}}, \bibnamefont{and}
  \bibinfo{author}{\bibfnamefont{S.-C.} \bibnamefont{Zhang}},
  \bibinfo{journal}{Phys. Rev. B} \textbf{\bibinfo{volume}{75}},
  \bibinfo{pages}{115103} (\bibinfo{year}{2007}).

\bibitem[{\citenamefont{{Wu}}(2009)}]{wu2009}
\bibinfo{author}{\bibfnamefont{C.}~\bibnamefont{{Wu}}}, \bibinfo{journal}{Mod.
  Phys. Lett. B} \textbf{\bibinfo{volume}{23}}, \bibinfo{pages}{1}
  (\bibinfo{year}{2009}), \eprint{0901.1415}.

\bibitem[{\citenamefont{Cai and Wu}(2011)}]{cai2011}
\bibinfo{author}{\bibfnamefont{Z.}~\bibnamefont{Cai}} \bibnamefont{and}
  \bibinfo{author}{\bibfnamefont{C.}~\bibnamefont{Wu}}, \bibinfo{journal}{Phys.
  Rev. A} \textbf{\bibinfo{volume}{84}}, \bibinfo{pages}{033635}
  (\bibinfo{year}{2011}).

\bibitem[{\citenamefont{Isacsson and Girvin}(2005)}]{isacsson2005}
\bibinfo{author}{\bibfnamefont{A.}~\bibnamefont{Isacsson}} \bibnamefont{and}
  \bibinfo{author}{\bibfnamefont{S.~M.} \bibnamefont{Girvin}},
  \bibinfo{journal}{Phys. Rev. A} \textbf{\bibinfo{volume}{72}},
  \bibinfo{pages}{053604} (\bibinfo{year}{2005}).

\bibitem[{\citenamefont{Liu and Wu}(2006)}]{liu2006}
\bibinfo{author}{\bibfnamefont{W.~V.} \bibnamefont{Liu}} \bibnamefont{and}
  \bibinfo{author}{\bibfnamefont{C.}~\bibnamefont{Wu}}, \bibinfo{journal}{Phys.
  Rev. A} \textbf{\bibinfo{volume}{74}}, \bibinfo{pages}{13607}
  (\bibinfo{year}{2006}).

\bibitem[{\citenamefont{Wu et~al.}(2006)\citenamefont{Wu, Liu, Moore, and
  Das~Sarma}}]{wu2006}
\bibinfo{author}{\bibfnamefont{C.}~\bibnamefont{Wu}},
  \bibinfo{author}{\bibfnamefont{W.~V.} \bibnamefont{Liu}},
  \bibinfo{author}{\bibfnamefont{J.~E.} \bibnamefont{Moore}}, \bibnamefont{and}
  \bibinfo{author}{\bibfnamefont{S.}~\bibnamefont{Das~Sarma}},
  \bibinfo{journal}{Phys. Rev. Lett.} \textbf{\bibinfo{volume}{97}},
  \bibinfo{pages}{190406} (\bibinfo{year}{2006}).

\bibitem[{\citenamefont{Kuklov}(2006)}]{kuklov2006}
\bibinfo{author}{\bibfnamefont{A.~B.} \bibnamefont{Kuklov}},
  \bibinfo{journal}{Phys. Rev. Lett.} \textbf{\bibinfo{volume}{97}},
  \bibinfo{pages}{110405} (\bibinfo{year}{2006}).

\bibitem[{\citenamefont{Stojanovi\ifmmode~\acute{c}\else \'{c}\fi{}
  et~al.}(2008)\citenamefont{Stojanovi\ifmmode~\acute{c}\else \'{c}\fi{}, Wu,
  Liu, and Das~Sarma}}]{stojanovic2008}
\bibinfo{author}{\bibfnamefont{V.~M.}
  \bibnamefont{Stojanovi\ifmmode~\acute{c}\else \'{c}\fi{}}},
  \bibinfo{author}{\bibfnamefont{C.}~\bibnamefont{Wu}},
  \bibinfo{author}{\bibfnamefont{W.~V.} \bibnamefont{Liu}}, \bibnamefont{and}
  \bibinfo{author}{\bibfnamefont{S.}~\bibnamefont{Das~Sarma}},
  \bibinfo{journal}{Phys. Rev. Lett.} \textbf{\bibinfo{volume}{101}},
  \bibinfo{pages}{125301} (\bibinfo{year}{2008}).

\bibitem[{\citenamefont{Zhou et~al.}(2011)\citenamefont{Zhou, Porto, and
  Das~Sarma}}]{zhou2011}
\bibinfo{author}{\bibfnamefont{Q.}~\bibnamefont{Zhou}},
  \bibinfo{author}{\bibfnamefont{J.~V.} \bibnamefont{Porto}}, \bibnamefont{and}
  \bibinfo{author}{\bibfnamefont{S.}~\bibnamefont{Das~Sarma}},
  \bibinfo{journal}{Phys. Rev. B} \textbf{\bibinfo{volume}{83}},
  \bibinfo{pages}{195106} (\bibinfo{year}{2011}).

\bibitem[{\citenamefont{M\"uller et~al.}(2007)\citenamefont{M\"uller,
  F\"olling, Widera, and Bloch}}]{mueller2007}
\bibinfo{author}{\bibfnamefont{T.}~\bibnamefont{M\"uller}},
  \bibinfo{author}{\bibfnamefont{S.}~\bibnamefont{F\"olling}},
  \bibinfo{author}{\bibfnamefont{A.}~\bibnamefont{Widera}}, \bibnamefont{and}
  \bibinfo{author}{\bibfnamefont{I.}~\bibnamefont{Bloch}},
  \bibinfo{journal}{Phys. Rev. Lett.} \textbf{\bibinfo{volume}{99}},
  \bibinfo{pages}{200405} (\bibinfo{year}{2007}).

\bibitem[{\citenamefont{{Wirth} et~al.}(2011)\citenamefont{{Wirth},
  {{\"O}lschl{\"a}ger}, and {Hemmerich}}}]{wirth2011}
\bibinfo{author}{\bibfnamefont{G.}~\bibnamefont{{Wirth}}},
  \bibinfo{author}{\bibfnamefont{M.}~\bibnamefont{{{\"O}lschl{\"a}ger}}},
  \bibnamefont{and}
  \bibinfo{author}{\bibfnamefont{A.}~\bibnamefont{{Hemmerich}}},
  \bibinfo{journal}{Nature Physics} \textbf{\bibinfo{volume}{7}},
  \bibinfo{pages}{147} (\bibinfo{year}{2011}).

\bibitem[{\citenamefont{\"Olschl\"ager
  et~al.}(2011)\citenamefont{\"Olschl\"ager, Wirth, and
  Hemmerich}}]{wirth2011a}
\bibinfo{author}{\bibfnamefont{M.}~\bibnamefont{\"Olschl\"ager}},
  \bibinfo{author}{\bibfnamefont{G.}~\bibnamefont{Wirth}}, \bibnamefont{and}
  \bibinfo{author}{\bibfnamefont{A.}~\bibnamefont{Hemmerich}},
  \bibinfo{journal}{Phys. Rev. Lett.} \textbf{\bibinfo{volume}{106}},
  \bibinfo{pages}{015302} (\bibinfo{year}{2011}).

\bibitem[{\citenamefont{{Kim} et~al.}(2011)\citenamefont{{Kim}, Kusudo, Wu,
  Masumoto, Schneider, H\"{o}ling, Kumada, Worschech, Forchel, and
  Yamamoto}}]{kim2011}
\bibinfo{author}{\bibfnamefont{N.~Y.} \bibnamefont{{Kim}}},
  \bibinfo{author}{\bibfnamefont{K.}~\bibnamefont{Kusudo}},
  \bibinfo{author}{\bibfnamefont{C.}~\bibnamefont{Wu}},
  \bibinfo{author}{\bibfnamefont{N.}~\bibnamefont{Masumoto}},
  \bibinfo{author}{\bibfnamefont{C.}~\bibnamefont{Schneider}},
  \bibinfo{author}{\bibfnamefont{S.}~\bibnamefont{H\"{o}ling}},
  \bibinfo{author}{\bibfnamefont{N.}~\bibnamefont{Kumada}},
  \bibinfo{author}{\bibfnamefont{L.}~\bibnamefont{Worschech}},
  \bibinfo{author}{\bibfnamefont{A.}~\bibnamefont{Forchel}}, \bibnamefont{and}
  \bibinfo{author}{\bibfnamefont{Y.}~\bibnamefont{Yamamoto}},
  \bibinfo{journal}{Nature Physics}  (\bibinfo{year}{2011}).

\bibitem[{\citenamefont{Feynman}(1972)}]{feynman1972}
\bibinfo{author}{\bibfnamefont{R.~P.} \bibnamefont{Feynman}},
  \emph{\bibinfo{title}{Statistical Mechanics, A Set of Lectures}}
  (\bibinfo{publisher}{Addison-Wesley Publishing Company, Boson},
  \bibinfo{year}{1972}).

\bibitem[{\citenamefont{Dalfovo et~al.}(1999)\citenamefont{Dalfovo, Giorgini,
  Pitaevskii, and Stringari}}]{pitaevskii1999}
\bibinfo{author}{\bibfnamefont{F.}~\bibnamefont{Dalfovo}},
  \bibinfo{author}{\bibfnamefont{S.}~\bibnamefont{Giorgini}},
  \bibinfo{author}{\bibfnamefont{L.~P.} \bibnamefont{Pitaevskii}},
  \bibnamefont{and}
  \bibinfo{author}{\bibfnamefont{S.}~\bibnamefont{Stringari}},
  \bibinfo{journal}{Rev. Mod. Phys.} \textbf{\bibinfo{volume}{71}},
  \bibinfo{pages}{463} (\bibinfo{year}{1999}).

\bibitem[{\citenamefont{Duan}(2006)}]{duan2006}
\bibinfo{author}{\bibfnamefont{L.-M.} \bibnamefont{Duan}},
  \bibinfo{journal}{Phys. Rev. Lett.} \textbf{\bibinfo{volume}{96}},
  \bibinfo{pages}{103201} (\bibinfo{year}{2006}).

\bibitem[{\citenamefont{Kitagawa et~al.}(2011)\citenamefont{Kitagawa, Aspect,
  Greiner, and Demler}}]{kitagawa2011}
\bibinfo{author}{\bibfnamefont{T.}~\bibnamefont{Kitagawa}},
  \bibinfo{author}{\bibfnamefont{A.}~\bibnamefont{Aspect}},
  \bibinfo{author}{\bibfnamefont{M.}~\bibnamefont{Greiner}}, \bibnamefont{and}
  \bibinfo{author}{\bibfnamefont{E.}~\bibnamefont{Demler}},
  \bibinfo{journal}{Phys. Rev. Lett.} \textbf{\bibinfo{volume}{106}},
  \bibinfo{pages}{115302} (\bibinfo{year}{2011}).

\bibitem[{\citenamefont{Yi et~al.}(2007)\citenamefont{Yi, Lin, and
  Duan}}]{yi2007}
\bibinfo{author}{\bibfnamefont{W.}~\bibnamefont{Yi}},
  \bibinfo{author}{\bibfnamefont{G.-D.} \bibnamefont{Lin}}, \bibnamefont{and}
  \bibinfo{author}{\bibfnamefont{L.-M.} \bibnamefont{Duan}},
  \bibinfo{journal}{Phys. Rev. A} \textbf{\bibinfo{volume}{76}},
  \bibinfo{pages}{031602} (\bibinfo{year}{2007}).

\bibitem[{\citenamefont{Lin et~al.}(2008)\citenamefont{Lin, Zhang, and
  Duan}}]{lin2008}
\bibinfo{author}{\bibfnamefont{G.-D.} \bibnamefont{Lin}},
  \bibinfo{author}{\bibfnamefont{W.}~\bibnamefont{Zhang}}, \bibnamefont{and}
  \bibinfo{author}{\bibfnamefont{L.-M.} \bibnamefont{Duan}},
  \bibinfo{journal}{Phys. Rev. A} \textbf{\bibinfo{volume}{77}},
  \bibinfo{pages}{043626} (\bibinfo{year}{2008}).

\bibitem[{\citenamefont{Saito et~al.}(2007)\citenamefont{Saito, Hulet, and
  Ueda}}]{saito2007}
\bibinfo{author}{\bibfnamefont{H.}~\bibnamefont{Saito}},
  \bibinfo{author}{\bibfnamefont{R.~G.} \bibnamefont{Hulet}}, \bibnamefont{and}
  \bibinfo{author}{\bibfnamefont{M.}~\bibnamefont{Ueda}},
  \bibinfo{journal}{Phys. Rev. A} \textbf{\bibinfo{volume}{76}},
  \bibinfo{pages}{053619} (\bibinfo{year}{2007}).

\end{thebibliography}

\end{document}